\documentclass[twocolumn,showpacs,preprintnumbers,amsmath,amssymb]{revtex4}
\usepackage{graphicx} 
\usepackage{dcolumn}

\begin{document}
\draft

\title{First-Principles Study of the Temperature-Pressure Phase
Diagram of BaTiO$_3$}

\author{Jorge \'I\~niguez and D. Vanderbilt}

\address{Department of Physics and Astronomy, Rutgers University,
Piscataway, New Jersey 08854-8019, USA}

\date{\today}

\begin{abstract}
We investigate the temperature-pressure phase diagram of BaTiO$_3$
using a first-principles effective-Hamiltonian approach. We find that the
zero-point motion of the ions affects the form of the phase diagram
dramatically. Specifically, when the zero-point fluctuations are included
in the calculations, all the polar (tetragonal, orthorhombic, and
rhombohedral) phases of BaTiO$_3$ survive down to 0~K, while only the
rhombohedral phase does otherwise. We provide a simple explanation for
this behavior. Our results confirm the essential correctness of the
phase diagram proposed by Ishidate {\it et al.}\ (Phys.  Rev. Lett.
{\bf 78}, 2397 (1997)).
\end{abstract}

\pacs{77.80.Bh, 64.70.-p, 61.50.Ks, 61.66.Hq}

\maketitle

BaTiO$_3$ is a paradigmatic example of a ferroelectric
material~\cite{ref:lin77}. Over the years it has been extensively
studied from both the experimental and theoretical points of view. It
is thus surprising to discover that its temperature-pressure phase
diagram remains very poorly investigated. Actually, even the
qualitative form of the phase diagram is still controversial.

\begin{figure}[b!]
\begin{center}
\includegraphics[width=6.5cm]{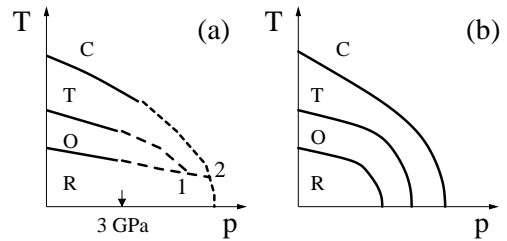}
\end{center}
\vskip 1mm
\caption{(a) Sketch of phase diagram of BaTiO$_3$ as discussed in
Refs.~\protect\cite{ref:sam66,ref:sam71,ref:guf76}. Phases are cubic
(C), tetragonal (T), orthorhombic (O), and rhombohedral (R).  Solid
and dashed lines represent measured data and suggested hypothetical
completion, respectively.  (Alternatively, critical points `1' and `2'
were suggested to coincide in Ref.~\protect\cite{ref:guf76}).  (b)
Sketch of phase diagram of BaTiO$_3$ as measured by Ishidate {\it et
al.}~\protect\cite{ref:ish97}.}
\label{fig:exppd}
\end{figure}

Figure~\ref{fig:exppd}a illustrates the results of early experimental
studies of the phase diagram of BaTiO$_3$~\cite{ref:sam66,ref:sam71},
which were confined to pressures up to $\sim$3~GPa only. In this
pressure range the system retains its zero-pressure transition
sequence with decreasing temperature. That is, it progresses from the
high-temeperature paraelectric cubic phase to ferroelectric
tetragonal, then ferroelectric orthorhombic, and finally ferroelectric
rhombohedral phases.  It was proposed that the diagram should be
completed as shown in Fig.~\ref{fig:exppd}a~\cite{ref:sam66} (or with
a modification in which the critical points labeled `1' and `2' meet
at a multicritical point~\cite{ref:guf76}). This kind of scenario has
been generally accepted ever since. On the other hand, in 1997
Ishidate {\it et al.}~\cite{ref:ish97} published the first (and, up to
now, the only) experimental study extending high enough in pressure to
reveal the actual form of the entire phase diagram.  Surprisingly,
these authors found that all the polar phases of BaTiO$_3$ survive
down to 0~K as sketched in Fig.~\ref{fig:exppd}b.

Ishidate {\it et al.}\ attributed this remarkable result to the
zero-point motion of the ions, which they argued should be significant
up to temperatures as high as 200~K. This explanation may appear
difficult to accept, given that no atom lighter than oxygen is present
in BaTiO$_3$. Indeed, it may be tempting to dismiss the phase diagram
of Ishidate {\it et al.}~on the basis of technical concerns. In
particular, it is not easy to find a pressure-transmitting medium that
remains fluid (i.e., isotropic) at the low temperature and high
pressure required for this study~\cite{ref:sam71}. This is a crucial
issue because anisotropies in the applied pressure could easily upset
the delicate balance of stabilities of the polar phases of BaTiO$_3$
and lead to incorrect results.  In view of these doubts, there is
clearly a pressing need for a fresh approach to this problem.

In this work, we use a first-principles effective-Hamiltonian
approach~\cite{ref:zho94,ref:zho95} to carry out a theoretical study
of the pressure-temperature phase diagram of
BaTiO$_3$. First-principles methods have been used extensively since
the early 1990s in many successful studies of ferroelectric
perovskites~\cite{ref:van97}. Our approach is well suited to the
present problem because it enables us (i) to calculate the
thermodynamic properties of BaTiO$_3$ in the presence of {\it
perfectly hydrostatic} pressures, and (ii) to switch on and off the
zero-point motion of the ions at will. Surprisingly, our results
corroborate the scenario proposed by Ishidate et
al.~\cite{ref:ish97}. We confirm that quantum fluctuations completely
change the high-pressure structure of the phase diagram and allow all
three ferroelectric phases to survive down to zero temperature, as
sketched in Fig.~\ref{fig:exppd}b. We discuss these results and
provide a simple explanation for the predicted form of the phase
diagram.

\vspace{1mm}

For this study we have made use of the effective-Hamiltonian approach
proposed for BaTiO$_3$ by Zhong {\it et
al.}~\cite{ref:zho94,ref:zho95}. The effective Hamiltonian is a
Taylor-series expansion of the potential energy of the system around a
high-symmetry phase, written in terms of a set of relevant degrees of
freedom. For BaTiO$_3$, the relevant variables are the local polar
modes (that add up to produce the spontaneous polarization of the
system) and the homogeneous strains, and the reference structure is
the paraelectric cubic phase. The parameters in this expansion are
obtained from first-principles density-functional
calculations~\cite{ref:explain-heff}. Zhong {\it et
al.}~\cite{ref:zho94} performed classical Monte Carlo (MC) simulations
on the basis of such an effective Hamiltonian and demonstrated that it
correctly reproduces the non-trivial phase transiton sequence of
BaTiO$_3$ along the {\it zero}-pressure isobar~\cite{ref:fn1}. Indeed,
after this initial achievement, the first-principles
effective-Hamiltonian method has been successfully applied over the
years to situations of increasing complexity~\cite{ref:bel}.

However, one should bear in mind that the quantitative accuracy of
this approach is still limited. The approximations involved in the
effective-Hamiltonian construction, including those related to the
first-principles methods used, result in some calculated quantities
(especially transition temperatures) that are not in very good
quantitative agreement with experiment. Of particular relevance for us
is the well-known understimation of the equilibrium volumes given by
the local-density approximation, which brings about a systematic error
in the location of our zero of pressure~\cite{ref:fn1}. For these
reasons, the results of the present calculations are to be regarded as
reliable only at the qualitative level.

{\sl Classical theory.---} We first calculated the phase diagram of
BaTiO$_3$ at a classical level by performing standard Monte Carlo
simulations for a number of temperatures and external hydrostatic
pressures. We simulated a $12\times 12\times 12$ supercell with
periodic boundary conditions, and typically did 30,000 MC sweeps to
thermalize the system and another 30,000 sweeps to calculate averages.
Our classical calculation is essentially a repeat of the one reported
in Fig.~4 of Ref.~\cite{ref:zho95}, except that we have taken special
pains to resolve the high-pressure part of the phase diagram as
carefully as possible.

\begin{figure}
%\begin{center}
\includegraphics[width=6cm,angle=-90]{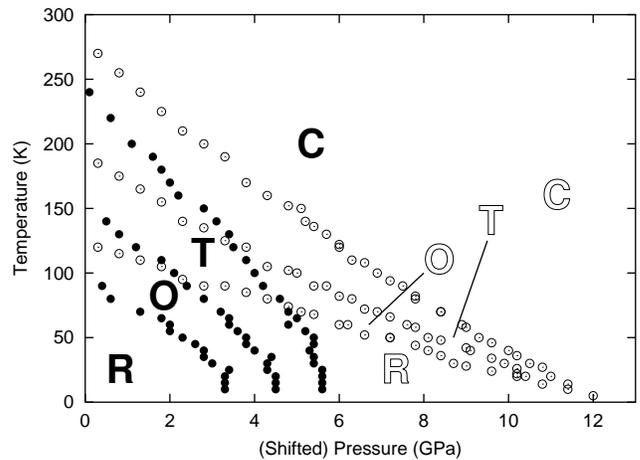}
%\end{center}
\vskip 2mm
\caption{Phase diagram of BaTiO$_3$ calculated at the classical level
(open circles and labels) and including quantum
fluctuations (solid circles and labels). Theoretical
pressures are corrected (shifted) following Zhong {\it et
al.}~\protect\cite{ref:fn1}.}
\label{fig:calpd}
\end{figure}

Our result, depicted with open circles in Fig.~\ref{fig:calpd}, is
topologically identical to the one shown in Fig.~\ref{fig:exppd}a.
The interesting action occurs in a small region of low temperature and
high pressure where the different phases meet.  In this region the
free-energy landscape of the system is extremely isotropic and it is
difficult to locate the phase boundaries precisely.  We can say with
some confidence that the rhombohedral and cubic phases meet along a
phase boundary that extends from about 12.5\,GPA at $T$=0\,K to about
11.5\,GPa at about 10\,K.  Whether all phases then meet at a
multicritical point, or whether there are two separate critical points
as illustrated in Fig.~\ref{fig:exppd}a, is difficult to decide
(although we tentatively favor the latter possibility).
A more reliable calculation adopting an approach such as that of
Ref.~\cite{ref:ini01}, which allows for a detailed exploration of the
free-energy landscape, would probably be needed to decide for certain.

The key conclusion we extract from our classical calculations is that,
provided the zero-point motion of the ions is not considered, only the
cubic and rhombohedral phases can survive down to zero temperature,
i.e., can be true ground states of the system.
To be certain that this conclusion is not an artifact of any
approximations made in connection with the effective-Hamiltonian
method, a careful check was carried out using zero-temperature
density-functional calculations directly~\cite{ref:inig02}. These
tests confirm the presence of a second-order transition directly from
the ferroelectric rhombohedral to the paraelectric cubic phase with
increasing pressure along the zero-temperature isotherm.

{\sl Quantum-mechanical theory.---} The ionic zero-point motion can be
included in our calculations by carrying out the thermodynamic
simulations using the path-integral quantum Monte Carlo (PI-QMC)
technique~\cite{ref:cep95} in place of the classical Monte Carlo. The
same effective Hamiltonian is used in both cases.  A preliminary study
of this kind, but limited to zero pressure and small Trotter numbers,
was initiated by Zhong and Vanderbilt~\cite{ref:zho96}, who showed
that the transition temperatures are indeed significantly affected by
the quantum-mechanical fluctuations. For instance, the rhombohedral to
orthorhombic transition, classically calculated to occur at 200~K, was
found to fall to 150~K. Not only is this effect quite large, but it is
also present at surprisingly high temperatures.

The technical details of the PI-QMC calculations are as follows. A
careful convergence analysis of simulations at 10~K led us to take a
Trotter number $P=64$ as a good compromise between accuracy and
computational feasibility. (Note that the size of the simulated system
is proportional to $P$.) For consistency, we kept the quantity $1/TP$,
which determines the degree of convergence of the PI-QMC results,
constant throughout the studied temperature range. In order to obtain
a thermalized configuration for a given $P$, we find it convenient to
increase $P$ from smaller values in a stepwise manner.  For example,
if our target is $P=12$, we consider $P = 1 \rightarrow 3 \rightarrow
6 \rightarrow 12$, feeding every new calculation with the thermalized
configuration obtained in the previous one. We typically performed
30,000 and 70,000 MC sweeps for thermalization and averages
respectively, using a $10\times 10\times 10$ supercell.  We checked
that these choices led to well-converged results.

Our result is depicted with filled circles in Fig.~\ref{fig:calpd}. We
find, in perfect qualitative agreement with Ishidate {\it et al.},
that all the polar phases of BaTiO$_3$ survive down to 0~K! Note the
dramatic {\it bending} of the transition lines, which pass from
following the classical law $T_c\propto (p_0 -p)$ at high temperatures
to following the quantum-mechanical law $T_c \propto (p_c -
p)^{1/2}$~\cite{ref:salje} at lower, but still relatively high,
temperatures. This is exactly the crossover that Ishidate
{\it et al.}~observed and which led them to attribute the occurrence of the
orthorhombic and tetragonal phases at 0~K to the zero-point motion of
the ions in the system. Our result clearly shows that such an
interpretation is correct.

\begin{figure}
\begin{center}
\includegraphics[width=5cm,angle=-90]{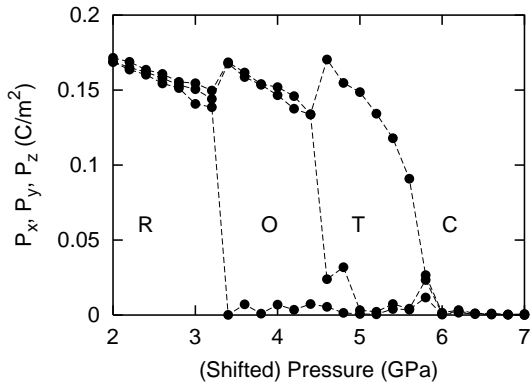}
\end{center}
\vskip 1mm
\caption{Calculated (quantum) phase transition sequence of BaTiO$_3$
along the 10~K isotherm, showing equilibrium polarization $(P_x, P_y,
P_z)$ as a function of pressure. Theoretical pressures are corrected
(shifted) following Zhong {\it et al.}~\protect\cite{ref:fn1}.}
\label{fig:10K}
\end{figure}

Figure~\ref{fig:10K} shows the calculated sequence of phase
transitions along the 10~K isotherm. It is apparent that, in spite of
the difficulties involved in our PI-QMC calculations, we are able to
identify the phase transitions unambiguously. The calculated pressure
range of stability of both the orthorhombic and tetragonal phases at
low temperatures is around 1~GPa, while in Ref.~\cite{ref:ish97}
ranges of about 0.6~GPa are reported. Also, we obtain a value of
approximately 6~GPa for the critical pressure $p_c$ at which
ferroelectricity disappears, in rough agreement with the value of
6.5~GPa obtained by Ishidate {\it et al}. Of course, given the
limitations of our method, this level of agreement may be partly
fortuitous.

{\sl Discussion.---} This quantum phase diagram can be rationalized in
the following way. It is natural to assume that in BaTiO$_3$ the
lattice-dynamical fluctuations, {\it either thermal or quantum
mechanical in character}, tend to favor the paraelectric cubic phase,
followed respectively by the tetragonal, orthorhombic, and finally
rhombohedral ferroelectric phases. On the other hand, we know from our
first-principles calculations on BaTiO$_3$ that the potential-energy
preference follows just the reverse order (for any pressure $p<p_c$).
We can thus view the phase transition sequence of this material along
the zero-pressure isobar as the result of the competition between
these two tendencies. Now let us turn to the case of the
zero-temperature isotherm, in which no thermal fluctuations are
present. At small pressures the rhombohedral phase is the ground state
because the potential energy contribution dominates over the ion
zero-point energy. However, as we compress the system, the
potential-energy differences between the different phases of BaTiO$_3$
decrease, as a direct consequence of the weakening of the
ferroelectric instability, and thus the {\it relative} importance of
the quantum-mechanical fluctuations grows accordingly.  Hence the
phase transition sequence can occur in a similar way along both the
zero-pressure isobar and the zero-temperature isotherm.

\begin{figure}
\begin{center}
\includegraphics[width=5cm,angle=-90]{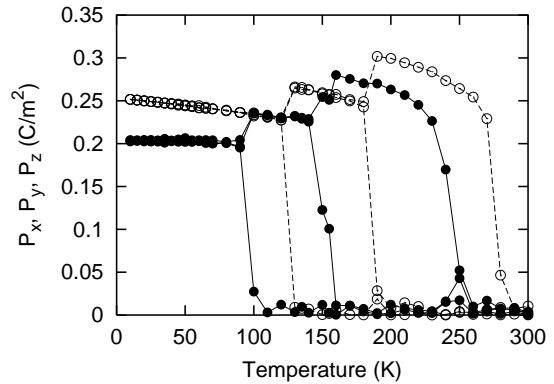}
\end{center}
\vskip 1mm
\caption{Phase transition sequence of BaTiO$_3$ along the zero-pressure
isobar, calculated at the classical (empty circles) and quantum-mechanical
(filled circles) level, showing equilibrium polarization $(P_x, P_y,
P_z)$ as a function of temperature. In both cases, the transition
sequence with decreasing temperature is
C$\rightarrow$T$\rightarrow$O$\rightarrow$R.}
\label{fig:qmsat}
\end{figure}

Figure~\ref{fig:qmsat} shows the calculated polarization along the
zero-pressure isobar, for both the classical and quantum cases. Classically
the polarization reaches 0~K with a finite slope, while the slope is
zero quantum mechanically. This is the expected quantum saturation of
the order parameter that has been discussed by Salje {\it et
al}.~\cite{ref:salje} in the context of structural phase
transitions. Following Ref.~\cite{ref:salje}, the quantum-saturation
effects in BaTiO$_3$ can be predicted to be significant up to several
hundred Kelvin, such high temperatures being a consequence of the
shallowness of the potential-energy wells associated with the
ferroelectric instabilities of the system. Indeed, in the pressure
range from about 3~GPa to about 6~GPa, this strong quantum
saturation can be regarded as inhibiting additional phase
transitions to lower potential-energy phases, thus allowing for the
occurrence of orthorhombic and tetragonal ground states.

It is important to note that the classical and quantum phase diagrams
depicted in Fig.~\ref{fig:calpd} cannot be related by the approximate
quantum-mechanical correction that is most common in the context of
Landau theories of phase transitions~\cite{ref:salje}. As shown in
Ref.~\cite{ref:ini01}, the classical free-energy landscape of
BaTiO$_3$ at low temperatures is described by a fourth-order expansion
in the polarization, sixth-order terms being identically zero at
0~K. On the other hand, some features of the quantum-mechanical phase
diagram at low temperatures, e.g. the existence of an orthorhombic
ground state, are strictly sixth-order in character. Since the usual
quantum-mechanical correction of the classical Landau potential only
involves a modified temperature dependence of the quadratic
term~\cite{ref:salje}, it could never account for such
features. Hence, the variety of accessible phases in BaTiO$_3$ results
in non-trivial quantum corrections and considerably hampers analytical
treatment of the problem, costly numerical solutions being required.

\vspace{1mm}

In summary, we have made use of the first-principles
effective-Hamiltonian method of Zhong {\it et al.}\ to study in detail
the temperature-pressure phase diagram of BaTiO$_3$. We have gone
beyond the usual approach and considered the zero-point motion of the
ions in our calculations by means of the path-integral quantum Monte
Carlo method. We find that the quantum fluctuations make a dramatic
difference with respect to the classical result. In the
quantum-mechanical case, all the polar phases of the system
(rhombohedral, orthorhombic, and tetragonal) survive down to 0~K,
while at the classical level only the rhombohedral phase does. Our
result is in essential agreement with the experimental work of
Ishidate {\it et al.}, thus giving strong support to the conclusions of
these authors.

We thank G. A. Samara for helpful discussions at the inception of this
project.  This work greatly benefited from previous technical
contributions of W. Zhong and A. Garc\'{\i}a.  The work was supported
by ONR grant N0014-97-1-0048, and made use of computational facilities
of the {\it Center for Piezoelectrics by Design}.

\end{document}